\documentclass{article}
\usepackage[a4paper, total={7.2in, 9.5in}]{geometry}
\usepackage[utf8]{inputenc}
\usepackage{blindtext}
\usepackage{ragged2e}
\usepackage{multicol}
\usepackage[font=sf]{caption}
\usepackage{fancyhdr}
\usepackage[pdftex]{hyperref}
\usepackage[dvipsnames]{xcolor}
\usepackage[most]{tcolorbox}
\usepackage{tabularx}
\usepackage{enumitem,kantlipsum}
\usepackage[font=sf]{caption}
\usepackage[font=sf]{floatrow}

\hypersetup{
breaklinks=true,   
colorlinks=true,   
}
\definecolor{textgreen}{RGB}{33, 143, 38}
\definecolor{headergreen}{RGB}{19, 88, 97}

\begin{document}
\sffamily

\noindent{\Huge Audible universe}

\vspace{0.2cm}{\large{\color{textgreen}\noindent Chris Harrison$^{1,\star}$, Anita Zanella$^{2,\dagger}$, Nic Bonne$^{3}$, Kate Meredith$^{4}$ and Nicolas Misdariis$^{5}$}}\\
{\em $^{1}$Newcastle University, Newcastle, UK. $^{2}$INAF, Padova, Italy. $^{3}$Portsmouth University, Portsmouth, UK. $^{4}$GLAS Education, Wisconsin, USA. $^{5}$STMS Ircam-CNRS-SU, Paris, France.}\\
\noindent Email: $^{\star}$\href{mailto:christopher.harrison@newcastle.ac.uk}{christopher.harrison@newcastle.ac.uk}, $^{\dagger}$\href{mailto:anita.zanella@inaf.it}{anita.zanella@inaf.it}
 
\vspace{0.6cm}\noindent{\large A multi-disciplinary team recently came together online to discuss the application of sonification in astronomy, focusing on the effective use of sound for scientific discovery and for improving accessibility to astronomy research and education.}

\begin{multicols*}{3}
\noindent Most of the matter in the Universe does not produce or absorb light, and even that which does is usually not visible to the human eye. Sophisticated technological and computational aids are required to turn astronomical data into visible images or into graphical data. However, challenging the idea that we should always use visualisations, interest in converting astronomical data into sound — through the process of ‘sonification’$^{1,2}$ — has risen in the past decade.  An example sonification of radio data from a fast radio burst is shown in Figure 1. For the purposes of this article, we refer to astronomy sonification projects more generally as any project that utilises sound, either directly or indirectly, to represent and/or communicate astronomical data or phenomena.

Consequently, for our workshop we brought together 55 experts working in previously disparate areas. Our goals were to start a multidisciplinary discussion on current attempts to sonify astronomical data, how to evaluate their efficacy and usefulness, and how to collaborate on an international standard framework for sonification of astronomical data by designing new tools and applications that also meet international accessibility criteria. 

Our online workshop, `Audible Universe', was hosted by the \href{https://www.lorentzcenter.nl/the-audible-universe.html}{Lorentz Centre}, from 31 August — 3 September 2021, after being advertised through their innovative \href{https://www.youtube.com/watch?v=nmnIF6rrNH8}{sonified poster}. 

\begin{figure*}
\centerline{\includegraphics[width=0.9\textwidth]{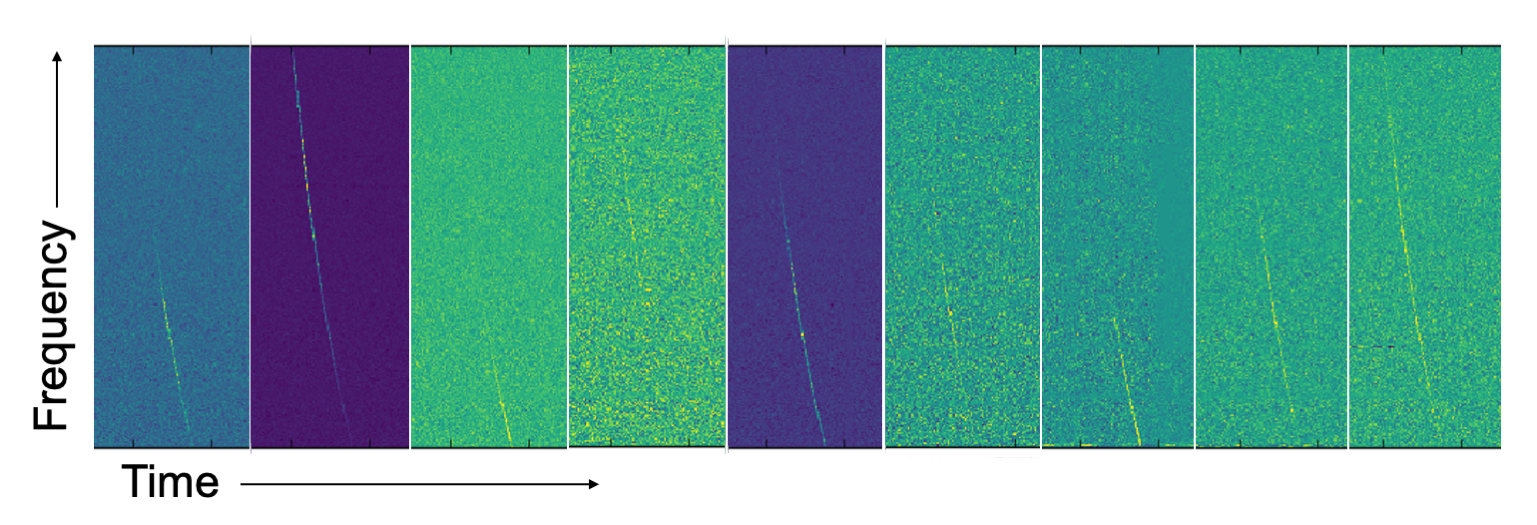}}
\RaggedRight{\footnotesize{\textbf{{\color{textgreen}Fig. 1.} Sonification of fast radio bursts.} The individual panels (left to right) are nine separate observations of a fast radio burst (FRB121102)$^{6}$. Radio frequency is plotted on the y-axis, and time on the x-axis. Each panel represents approximately the central half a second of data, centred on the burst time. The \href{https://www.youtube.com/watch?v=i3x0sBCQ_c8}{video and sound version of this figure} moves from left-to-right through each of these data points and sonifies them, associating high radio frequency with higher pitch and low radio frequency with lower pitch. This figure was adapted from the original data (ref. $^{7}$). Movie/Sonification Credit: Casey Law (ref. $^{8}$).}

}
\end{figure*}

\vspace{0.3cm}\noindent\textbf{Bringing together cross-disciplinary experts}

\noindent We invited experts in sound design, sonification, sound perception, and computer-human interactions to this workshop. To avoid confusion, we broadly categorised experts in these research fields as ‘sound people’ and those working primarily in astronomy research or education as ‘star people’. The star people gave three pre-recorded review talks on: (1) existing astronomy sonification projects related to education; (2) sonification for research using one-dimensional and time series data; and (3) sonification for research using two-dimensional (e.g. images) and multidimensional data (e.g. integral field spectroscopy). These are available on our \href{https://www.youtube.com/channel/UC_pFRNR2M4_6UA83dBb7nfA}{YouTube channel}.

These talks highlighted an impressive array of potential applications and inspiring anecdotal evidence on how well these tools have been received (particularly in the visually impaired specialised education setting). However, some areas of concern raised during the following discussion included very strong similarities between many of the tools (potentially a case of “re-inventing the wheel”), a lack of evidence-based approaches to the design process, and a lack of scientifically-driven experiments proving whether the projects had achieved their goals. These topics featured heavily in the following discussion during the workshop and are touched upon in a review article by Anita Zanella et al (Nature Astronomy, in press)$^{9}$.

The sound people gave pre-recorded introductory lectures covering the topics of: (1) sound perception and design; (2) data sonification; and (3) perceptual evaluation methods for sound-producing objects (with specific questions related to astronomy). These talks, available on our \href{https://www.youtube.com/channel/UC_pFRNR2M4_6UA83dBb7nfA}{YouTube channel}, highlighted some well-known aspects in the sound people’s community that had not necessarily reached the star people. They clarified terminology and methods commonly used for sonification (also see ref. $^{3}$). They further reported on commonly accepted parameter mapping choices -- highlighting, for example, that timbre is better for classification, whereas pitch is better for comparison of similar datasets. A summary of standard accepted evaluation processes for different types of activities was also provided, such as identification or categorisation of phenomena, scaling or rating given properties, sorting groups of objects, and working with multi-modal objects. More information about this will be provided in Nicolas Misdariis et al. (in preparation). 

\vspace{0.3cm}\noindent\textbf{Accessibility considerations in astronomy education and research}

\noindent On the first day of live sessions, we discussed accessibility considerations, particularly by visually impaired audiences and researchers with potential barriers to accessing astronomy. This discussion took the form of two Q\&A panel sessions during which many topics were covered, including the need to raise general awareness that it is possible to be blind and be a scientist. Fundamental challenges to research were mentioned, such as accessing the mathematical and graphical content of publications and the need to include accessibility into the design of new facilities and tools. For example, modern graphical user interfaces are usually less accessible than the more traditional command-line approach of operating software. These topics are presented in more detail in Jake Noel-Storr and Michele Willebrands (Nature Astronomy, in press)$^{10}$.

\vspace{0.3cm}\noindent\textbf{Evaluation and analysis of sonification projects}

\noindent In preparing for this workshop, the organisers noted a deficit in formal evaluation (or ``analysis”) of sonification tools and resources that had been produced by the astronomical community. Therefore, the goal of the second day of live sessions was to learn from the sound people who had expertise in scientific approaches to the evaluation of sound perception. Participants were split into small working groups to outline an evaluation strategy for five different existing astronomy-focused sonification tools (AstreOS, Astronify, SonoUno, A4BD and Afterglow Access). The developers had pre-recorded tutorial videos on their respective tool, and most of these are available on our \href{https://www.youtube.com/channel/UC_pFRNR2M4_6UA83dBb7nfA}{YouTube channel}. 

During the workshop, certain project leaders expressed that they had undertaken some informal evaluation. However, this evaluation was mostly carried out without following standard evaluation procedures and, crucially, is unpublished. A couple of published exceptions come from the thesis of Wanda Diaz-Merced$^{4}$ who demonstrates manners to optimise the simultaneous use of visuals and sonification and the work by Beatriz Garcia et al.$^{5}$ who demonstrates that most sonification software at the time failed to meet some basic international accessibility standards.

The working groups focused on the evaluation of different features key to the attainment of astronomical data that could easily be exported to other scientific fields: shape discrimination and recognition; object identification and classification; and object detection. They have evinced, prompted also by the pre-recorded talk on perceptual evaluation methods, how different evaluation techniques could be used depending on the goal of the sonification. For example, shape recognition can be assessed by asking the users to draw what they are listening to and the efficacy of sonification for object detection can be tested by measuring the detection speed. The working groups also highlighted the importance of evaluating well-defined aspects of tools and sonifications and focusing on one question at a time. Indeed, asking the users to perform an identification or a discrimination task, but not both at the same time, increases accuracy.

Two main approaches to evaluation were suggested: either asking large numbers of users to perform a given task and statistically evaluate their responses or working with individual users to get more specific feedback. The first approach is preferable to obtain statistical results on the general usefulness, efficacy, or usability of a tool. This option might most regularly be applied for non-expert audiences. The second approach might be useful in the design phase of a tool or to tailor it toward the specific needs of specific users.

Finally, several groups discussed the importance of evaluating the usefulness, and relative levels of satisfaction/enjoyment of sonification against other sensory approaches such as visualization and haptic representations. 

\vspace{0.3cm}\noindent\textbf{Design process for future projects}

\noindent The third day of the live sessions was dedicated to improving the design process and to learn how to more effectively design future applications. To facilitate progress in this area, participants were split into working groups to outline a design strategy for possible astronomy applications.

Some working groups focused on general applications that could easily be extended to other research fields, such as the sonification of three-dimensional cubes and the creation of physical or virtual sonic landscapes (for exhibitions or the exploration of datasets, for example). Other groups worked instead on specific applications such as: (1) a sound-based astronomical observations planner that could be used by professional astronomers at the telescope or by university students for education; (2) a citizen-science project to find exoplanet transits using sound; and (3) the development of a sound-based planetarium show.

A recurring discussion focused on how to give the audience an overview of the data while also conveying more details. This problem related to processing local and global information seems well-known by the community of sound people, and they suggested offering different layers of information to the users, that is, a more holistic event in one layer and more detailed information in another. It could also be possible to let the user modify the sound properties and customize the sonification in some of these layers. The discussion led to the agreement that both standardisation through sets of guidelines and customisation are important. Standardisation makes sonification choices recognizable whilst customization and interactivity actively engage the listeners and also ensures that perceptual differences among individuals are accounted for. Ensuring the correct balance between these factors will be important in the participant’s shared goal to have wide-spread acceptance, in the astronomy community, of sonification as an education and research tool. 

\vspace{0.3cm}\noindent\textbf{Future outlook}

\noindent The cross-disciplinary nature of the meeting is one of its strongest points, and most of the participants are willing to continue collaborating with each other on specific projects borne during the meeting. Recurrent discussion topics that could not be addressed in depth due to the lack of time but are worth delving into in future follow-up meetings include: (1) the need to train the public and include sonification in the school curricula from an early age; (2) the ways to overcome researcher skepticism to make sonification a mainstream tool; (3) the need (or not) to pay attention to the aesthetics of sonifications; (4) the ways emotions enter sonification; (5) the need to account for multicultural approaches to sonification;  and (6) the use of gamification to test and disseminate sonification. We believe that this is just the beginning of a fruitful exploration where different communities can proceed hand-in-hand to learn from each other on how best to apply sonification into astronomy research and education.

\vspace{0.5cm}\noindent{\color{textgreen}References}
\vspace{-0.15cm}

\begin{enumerate}[leftmargin=*]
  \setlength{\itemsep}{1pt}
  \setlength{\parskip}{0pt}
  \setlength{\parsep}{0pt}
\footnotesize{
\item Kramer, G. et al. {\em Sonification Report: Status of the Field and Research Agenda.} (Department of Psychology, University of Nebraska, 2010); \url{https://digitalcommons.unl.edu/psychfacpub/444}.
\item Hermann, T., Hunt, A., \& Neuhoff, J. G. {\em The Sonification Handbook} (Logos Publishing House, 2011).
\item Dubus, G., \& Bresin, R. {\em PloS One 8}, e82491 (2013).
\item Diaz-Merced, W. {\em Sound for the exploration of space physics data.} PhD thesis, Univ. of Glasgow (2013).
\item Garcia, B., Diaz-Merced, W., Casado, J. \& Cancio, A. {\em EPJ Web Conf.} \textbf{200}, 01013 (2019).
\item Chatterjee, S. et al. {\em Nature} \textbf{541}, 58 (2017).
\item Law, C. The Sound of Fast Radio Burst FRB 121102, {\em Harvard Dataverse}, \url{https://doi.org/10.7910/DVN/QSWJE6} (2016).
\item Law, C. Analysis of Very Large Array data toward FRB 121102, {\em Harvard Dataverse}, \url{https://doi.org/10.7910/DVN/ZKESD4} (2017). 
\item Zanella, A. et al. Sonification and sound design for astronomy research, education and public engagement, {\em Nature Astronomy} (2022), DOI: 10.1038/s41550-022-01721-z.
\item Noel-Storr, J., \& Willebrands, M., Accessibility in Astronomy for the Visually Impaired, {\em Nature Astronomy} (2022), DOI: 10.1038/s41550-022-01691-2.
}
\end{enumerate}

\vspace{0.3cm}\noindent{\color{textgreen}Acknowledgements}

\noindent{\small We thank the Lorentz Centre for hosting the workshop (online) and their valuable assistance in preparing and running the workshop. We thank all of the participants of the workshop, who all provided important contributions to the fruitful discussions.}

\vspace{0.3cm}\noindent{\color{textgreen}Competing interests}

\noindent{\small The authors declare no competing interests.}

\end{multicols*}

\end{document}